\documentclass[aps,groupedaddress]{revtex4-2}

\usepackage{amsfonts}
\usepackage{amsmath}
\usepackage{amssymb}
\usepackage{braket}
\usepackage[margin=1in]{geometry}
\usepackage{graphicx}
\usepackage{hyperref}
\usepackage{natbib}
\usepackage{setspace}
\usepackage{xspace}

\hypersetup{colorlinks=true }

\newcommand{\bb}[1]{\ensuremath{\mathbb{#1}}}
\newcommand{\mb}[1]{\ensuremath{\mathbf{#1}}}
\newcommand{\qi}{\mb{i}\xspace}
\newcommand{\qj}{\mb{j}\xspace}
\newcommand{\qk}{\mb{k}\xspace}
\newcommand{\ov}{\overline}

\begin{document}


\title{Nonlinear Addition of Qubit States Using Entangled Quaternionic Powers of Single-Qubit Gates}


\author{Dominic Widdows}
\email{widdows@ionq.com}
\affiliation{IonQ}


\date{\today}

\begin{abstract}
This paper presents a novel way to use the algebra of unit quaternions to express
arbitrary roots or fractional powers of single-qubit gates, and to use such fractional 
powers as generators for algebras that combine these fractional input signals, 
behaving as a kind of nonlinear addition.

The method works by connecting several well-known equivalences.
The group of all single-qubit gates is $U(2)$, the unitary 
transformations of $\mb{C}^2$. Using an appropriate phase multiplier, every element of $U(2)$ can be mapped to a corresponding
element of $SU(2)$ with unit determinant, whose quantum mechanical behavior is identical. The group $SU(2)$ is isomorphic to the group
of unit quaternions. Powers and roots of unit quaternions can be constructed by extending de Moivre's theorem
for roots of complex numbers to the quaternions by selecting a preferred `square root of -1'. Using this chain of equivalences,
for any single-qubit gate $A$ and real exponent $k$, a gate $B$ can be predictably constructed so that $B^k = A$. 

Different fractions generated in this way can be combined by connecting the individually rotated qubits to a 
common `sum' qubit using entangling 2-qubit CNOT gates.
Some of the simplest such algebras are explored --- those generated by roots of the quaternion \qk (which corresponds to
and $X$-rotation of the Bloch sphere), the quaternion $\frac{\sqrt{2}}{2}(\qi + \qk)$ (which corresponds to the Hadamard gate), and a mixture of these.

One of the goals of this research is to develop quantum versions of classical components such as 
the classifier ensembles and activation functions used in machine learning and artificial intelligence.
An example application for text classification is presented, which uses fractional rotation gates to represent 
classifier weights, and classifies new input by using 2-qubit CNOT gates to collect the appropriate classifier weights
in a topic-scoring qubit. 

\end{abstract}

\maketitle

\section{Introduction}

Transformations in quantum theory are expected to be continuous and reversible \cite{hardy2001quantum}. For any 
transformation $A$, we should therefore be able to find some partial transformation $B$ such that performing this transformation
$k$ times recovers $A$, which is expressed using the equation $B^k = A$. The problem of finding some suitable $B$
can be solved for general matrices $A$ and real powers 
$k$, but general solutions rely on detailed machinery such as Schur decomposition and Pad\'e approximation \cite{higham2011schur}.

For operations on single qubits, the group of possible transformations is much smaller and simpler than the general case. 
The qubit state is represented by a normalized vector $\alpha\ket{0} + \beta\ket{1}$ with $|\alpha|^2 + |\beta|^2 = 1$, and the transformations on
this state are given by the unitary group $U(2)$, the subgroup of $GL(2, \mb{C})$ that preserves this invariant. This observation leads to
the small set of single-qubit transformations or `gates' familiar in quantum computing \cite[\S1.3.1, \S4.2]{nielsen2002quantum}.
It is well-known that the space of single-qubit gates corresponds to rotations of the Bloch sphere $S^2$, and that this
space is 3-dimensional --- for example, a rotation of $S^2$ can be specified by giving 
a direction for the rotation axis (2 coordinates, e.g. latitude and longitude), and a rotation angle (1 more coordinate).
A common way to describe this space is to use three {\it Euler angles}, which are the parameters $\theta$, $\phi$ and $\lambda$
in the `universal' or {\it U}-gate \cite*[\S1.4]{qiskit2021textbook}. Euler angles are well-established in many fields and quite intuitive to understand,
but also have well-known drawbacks including coordinate singularities and gimbal lock \cite{diebel2006representing}.
This makes composing and decomposing sequences of rotations using Euler angles cumbersome, which has encouraged
the adoption of mathematical alternatives including quaternions for applications in computer science including graphics
\cite{mukundan2002quaternions}, aerospace, and virtual reality \cite{kuipers1999quaternions}.

This paper demonstrates that the use of quaternions instead of Euler angles enables
fractional powers of single-qubit gates to be constructed easily and explicitly, and that their behavior
when composed into larger circuits can be predicted simply and effectively. 
The method relies on connecting the following observations:

\begin{itemize}
\item Every gate represented by a unitary matrix $A$ can also be represented by a corresponding 
{\it special} unitary matrix (a matrix with unit determinant) by multiplying the matrix by the phase factor $\det(A)^{-\frac{1}{n}}$, where
$n$ is the number of (complex) dimensions which in the case of single-qubit gates is $n=2$.
\item The special unitary group $SU(2)$ is isomorphic to the group of unit quaternions, and so 
single-qubit gates can be represented by unit quaternions \citep{wharton2015unit}.
\item Fractional real roots of quaternions can be obtained using the quaternionic version of the de Moivre formula 
$(\cos(\theta) + i \sin(\theta))^n =  \cos(n\theta) + i \sin(n \theta)$ \citep{niven1942roots}. 
\item The quaternionic root found in this way can be mapped back to the corresponding special unitary matrix, which 
can be multiplied by a real root of the phase factor if we want to recover the original unitary matrix exactly.
\item These observations are combined to give an explicit formula for any real power of a single-qubit gate.
\end{itemize}

Once these fractional operations are available, they can be combined in simple `adder circuits' that use
2-qubit CNOT gates to connect
each of the fractionally rotated qubits to a common `sum' qubit that collects the various contributions.
The method is applied in a demonstration problem in text classification. 
Small fractional rotations are repeated to count the number of times a word is seen with a particular topic in training.
When classifying a new phrase, these rotation signals are accumulated by connecting the corresponding 
word-counting qubits to a scoring qubit using CNOT gates. 
When two such fractional rotations through angles $\theta$ and $\varphi$ are combined into the same scoring qubit, this gives 
an outcome probability that is a function $F$ of $\theta$ and $\varphi$.
This behavior depends on which quaternion generators were chosen --- for example, the quaternion $q = \cos(\theta) + \qk\sin(\theta)$ 
corresponding to the unitary gate $R_x(\theta)$ gives rise to a scoring function whose probability of giving an
output  $\ket{1}$ from the input angles $\theta$ and $\varphi$ is given by $\cos^2(\theta/2)\sin^2(\varphi/2) + \sin^2(\theta/2)\cos^2(\varphi/2)$. 
This application was chosen partly for its relevance to quantum machine learning, one of whose challenges is
to find suitable nonlinear activation functions \cite[\S 5.4.1]{schuld2021machine}.

The rest of this paper is arranged as follows. Section \ref{quaternion-sec} recalls the basic properties of quaternions, just enough to 
explain the passage from single-qubit unitary operators, to quaternions, to fractional quaternion powers, back to fractional powers of unitary operators.
Section \ref{addition-sec} shows how two of these fractional rotations can be wired together in a 3-qubit circuit that combines
the contributions from each input fraction, and compares three such functions arising from different quaternion generators.
Section \ref{classification-sec} demonstrates the successful application of these components in a simple 
text classification problem, where the accurate results also demonstrate that inaccuracies in gate-level qubit operations do
not always lead to application-level errors. Finally, Section \ref{related-sec} relates the work in this paper to known alternatives,
and proposes further work expanding on common areas.

\section{Quaternion Powers for Fractional Single-Qubit Gates}
\label{quaternion-sec}

It is assumed that the reader is familiar with quantum gates and unitary matrices (otherwise see \cite[\S2.1]{nielsen2002quantum}). Quaternions
are less ubiquitous in quantum theory, so their important properties are recalled here. 

Quaternions were discovered in 1843 by William Rowan Hamilton (1805--1865), who worked closely with Arthur Caley (1821--1895) on the foundations of linear algebra, 
introduced the terms `vector', `associative', `commutative', and `distributive' \cite{hamilton1847quaternions}, 
and also developed Hamiltonian mechanics.
The quaternions \bb{H} are a 4-dimensional real algebra generated by the identity element 1 and the symbols \qi, \qj, and \qk, multiplied according to the {\it quaternion relations}
\[ \qi\qj = -\qj\qi = \qk \quad \quad \qj\qk = -\qk\qj =  \qi \quad \quad \qk\qi = -\qi\qk = \qj \quad \quad \qi^2 = \qj^2 = \qk^2 = -1 \]
and the distributive law. (Note the similarity with the vector cross product in 3-dimensional coordinate geometry, 
and the algebra of Pauli matrices in quantum mechanics.) 
The quaternion algebra is not commutative, though it does obey the associative law (the investigation of 
these properties of quaternions led Hamilton to invent those terms).
The quaternions are a division algebra (an algebra with the property that $ab= 0$ implies that $a= 0$ or $b= 0$).

Each of the generators \qi, \qj, \qk behaves as a `square root of $-1$', and more generally, so does every imaginary quaternion $a\qi +  b\qj + c\qk$ with $a^2 + b^2 + c^2 = 1$.
Identifying such an $a\qi +  b\qj + c\qk$ with the imaginary number $i=\sqrt{-1}$ gives an embedding of the complex numbers $\bb{C} \hookrightarrow \bb{H}$. 
(To avoid confusion in this paper, the symbol $i$ will be used for the complex square root of $-1$, and the boldface symbol \qi will be used for the imaginary quaternion.)
The set of quaternionic square roots of $-1$ can be identified with the 2-sphere $S^2$. A quaternion $q = q_0 + q_1\qi + q_2\qj + q_3\qk$ can be decomposed into its
{\it real part} $q_0$ and its {\it imaginary part} $q_1\qi + q_2\qj + q_3\qk$, which Hamilton and his immediate successors referred to as the {\it scalar} and {\it vector} parts of the quaternion.

As with complex numbers, the {\it conjugate} of a quaternion $q = q_0 + q_1\qi + q_2\qj + q_3\qk$ is defined as $\bar{q} = q_0 - q_1\qi - q_2\qj - q_3\qk$, the {\it norm} of a 
quaternion is given by the formula $|q| = q\bar{q}$, and the {\it inverse} is given by $q^{-1} = \bar{q}/|q|$. 
Quaternions have a polar decomposition $q = |q| (\cos(\theta) + \gamma\sin(\theta)) = |q| e^{\gamma \theta}$, where $\gamma$ is one of the 2-sphere of unit imaginary quaternions. 
The {\it unit quaternions} are those with unit norm. A unit quaternion $q$ can therefore be written as $q = \cos(\theta) + \gamma\sin(\theta)$, where $\gamma$ again 
is an imaginary unit quaternion $\gamma\in \bb{H}$ such that $\gamma^2 = -1$. The group of unit quaternions is often called $Sp(1)$, because it is the first in the
series of symplectic groups \cite[\S7.2]{fulton2013representation}, though the relationship between symplectic geometry and quantum mechanics is not needed in this paper.

See \citet[Ch 1]{widdows2000quaternion} for more explanation of quaternion terminology, simple constructions using quaternions such as 3d and 4d rotations, 
and isomorphisms with various real and complex matrix groups and manifolds. The important relationship between $SU(2)$ and the unit quaternions is explained below.

\subsection{Real Fractional Roots of Quaternions}

Crucially for this work, the de Moivre formula 
\[(\cos(\theta) + i \sin(\theta))^n =  \cos(n\theta) + i \sin(n \theta)\] 
applies unchanged to the quaternionic setting \citep{niven1942roots}. That is, for a unit quaternion $q = \cos(\theta) + \gamma\sin(\theta)$ with $\gamma^2 = -1$, 
\begin{equation}q^n = \cos(n\theta) + \gamma\sin(n\theta).
\label{quat_polar}
\end{equation}
This adapts immediately to fractional powers: if $p = q^n = \cos(n\theta) + \gamma\sin(n\theta)$ 
it follows that $q = p^{1/n}$, so setting $\varphi = n \theta$, it follows that if $p = \cos(\varphi) + \gamma(\sin \varphi)$, then 
\begin{equation} p^{1/n} = \cos(\varphi / n) + \gamma(\sin \varphi / n). 
\label{qdemoivre}
\end{equation}
In general, quaternion roots, like complex roots, are usually not uniquely defined. Any multiple of $2\pi$ can be added to the angle $\varphi$ without 
changing the quaternion $p$, so typically the equation 
$q = p^{\frac{1}{n}}$ has $n$ roots if $n$ is an integer, infinitely many roots if $n$ is irrational, with special cases for when $p$ is a real number \citep{niven1942roots}.

\subsection{Single-Qubit Gates and Quaternions}
\label{gate_quat_sec}

Recall that single-qubit gates are represented by operators in the group $U(2)$, which can be characterized as those matrices $\{A \in GL(2, \mb{C}) : A\bar{A}^T = I\}$, where $\bar{A}$ is 
the complex conjugate matrix, $A^T$ the transpose, and $I$ is the identity matrix. All unitary matrices have determinants in the complex unit circle group $U(1)$, 
whereas special unitary matrices are those with determinant equal to 1. The group of such $2\times2$ matrices is called $SU(2)$.

The group of unit quaternions under multiplication is isomorphic to the multiplicative group $SU(2)$ 
with a correspondence given by

\begin{equation}  q = a + b\qi + c\qj + d\qk \longleftrightarrow A = \begin{pmatrix} a + bi & c + di \\ -c + di & a - bi\end{pmatrix}, \mathrm{where}\ a^2 + b^2 + c^2 + d^2 = 1.
\label{quat_to_su}
\end{equation}
It is easy to check that the condition $a^2 + b^2 + c^2 + d^2 = 1$ guarantees that $\det(A) = 1$ and $A\bar{A}^T = I$, so $A$ is indeed an element of $SU(2)$.

The Pauli matrices \cite[\S2.1.3]{nielsen2002quantum} are obtained from this construction as follows:

\begin{equation*}
\begin{aligned}
1 \in \mb{H} \longleftrightarrow
\begin{bmatrix} 
1 & 0 \\
0 & 1
\end{bmatrix}
\equiv \sigma_0 \equiv I
&&&&&&
\qi \in \mb{H} \longleftrightarrow
\begin{bmatrix} 
i & 0 \\
0 & -i
\end{bmatrix}
\equiv i \sigma_z \equiv iZ
\\
\qj \in \mb{H} \longleftrightarrow
\begin{bmatrix} 
0 & 1 \\
-1 & 0
\end{bmatrix}
\equiv -i \sigma_y \equiv -iY
&&&&&&
\qk \in \mb{H} \longleftrightarrow
\begin{bmatrix} 
0 & i \\
i & 0
\end{bmatrix}
\equiv i \sigma_z \equiv iX
\end{aligned} 
\end{equation*}

Note the multiplication by $\pm i$ to recover the Pauli matrices $\sigma_x$, $\sigma_y$, $\sigma_z$ in their most standard form.
This is allowed because multiplying by any global phase factor $e^{i\theta}$ makes no physical difference, and is necessary to be able
to find a quaternion corresponding to every unitary transformation. This strategy generalizes to other unitary matrices:
since the group $U(2)$ is more general than the normal subgroup $SU(2)$, not all unitary matrices representing single-qubit quantum gates 
have corresponding quaternions given by equation \ref{quat_to_su}, but all unitary matrices are phase-equivalent to
a special unitary matrix for which the quaternion correspondence works. An appropriate special unitary matrix can be found as follows.
For $A \in U(n)$, suppose that $\det(A) = e^{i \theta}$. It follows that $\det(e^{-\frac{i \theta}{n}} A) = 1$, and that this operator produces the same physical state as $A$. 
Thus we define a mapping $A \rightarrow \det(A)^{-\frac{1}{n}} A$, which assigns to each unitary matrix a special unitary matrix with the same physical effect. This mapping is
in fact a fibration mapping $U(n) \mapsto SU(n)$. 

The path for finding fractional real powers of a single-qubit quantum logic gate $A$ now takes shape. To find a gate $B$ such that $B^k = A$ for some power $k$, we follow the following steps:

\begin{itemize}
\item Take any single-qubit gate / unitary matrix $A\in U(2)$.
\item Map it to a representative special unitary matrix in $A^\prime \in SU(2)$ using the formula $A^\prime = \det(A)^{-\frac{1}{2}}A$.
\item Map $A^\prime \in SU(2)$ to a unit quaternion $q$ using the correspondence in equation \ref{quat_to_su}.
\item Take the $k^{th}$ root $q^\frac{1}{k}$ using equation~\ref{qdemoivre}.
\item Map this back to a special unitary matrix $B^\prime$, again using equation \ref{quat_to_su}. 
From the isomorphism between unit quaternions and $SU(2)$, it follows that $(B^\prime)^k = A^\prime$.
\item (Optional) To recover the matrix $A$ as well as the phase-equivalent $A^\prime$, multiply $B^\prime$ by a phase factor $\det(A)^{\frac{1}{2k}}$,
and verify that if $B = \det(A)^{\frac{1}{2k}} B^\prime$, then $B^k = A$.
\end{itemize} 

For dimension $n=2$, note that each unitary matrix $A$ has two possible 
special unitary images under the fibration mapping $A \rightarrow \det(A)^{-\frac{1}{2}} A$, because if $\det(A) = 1$ then $\det(-A) = 1$.
This corresponds to the multiplicity of roots in the expression $\det(A)^{-\frac{1}{2}}$. In algebraic terms, $U(2)$ is not a direct product $SU(2)\oplus U(1)$, though there is a short exact sequence
$1 \rightarrow SU(2) \rightarrow U(2) \rightarrow U(1) \rightarrow 1$. In terms of the correspondence with unit quaternions, this amounts to the observation that the unit quaternions
$q$ and $-q$ produce the same gate operation.

\section{Component Application --- Weighted Addition}
\label{addition-sec}

The original motivation for this work was to create a kind of quantum addition for fractional states, for use as a component in
 machine learning operations such as building classification / decision processes, and for activation functions in neural networks.
 This section demonstrates this compositional behavior.

\begin{figure}[ht]
\caption{An Example Addition Circuit}
\label{adder_circuit}
\bigskip
\centering
\includegraphics[width=2.2in]{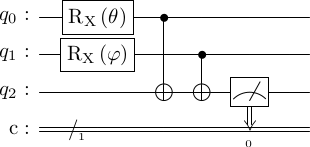}
\end{figure}

The basic idea is that fractional rotations in various qubits can be combined by entangling each rotation qubit with a common
target or scoring qubit using CNOT gates.
An example circuit implementation is shown in Figure \ref{adder_circuit}. There are two `summand' qubits $q_0$ and $q_1$, each of these has been put through
a single-qubit fractional gate transformation, and these qubits are each connected to a third `sum' qubit $q_2$. The result in the sum qubit is then measured.
In this example, the rotation $\mathrm{R_X}(\theta)$ was used, which corresponds to the quaternion $\cos(\theta) + \qk \sin(\theta)$. 
This example is particularly well-behaved, both algebraically and for using standard gate notation, though other quaternion generators
can be used.

In the rest of this section, the following generators were used:

\begin{equation}
\label{x_frac_eqn}
\qk^\alpha \longleftrightarrow
\begin{bmatrix} 
\cos(\frac{\pi\alpha}{2}) & i \sin(\frac{\pi\alpha}{2}) \\
i \sin(\frac{\pi\alpha}{2}) & \cos(\frac{\pi\alpha}{2})
\end{bmatrix}
\longleftrightarrow X^\alpha
\end{equation}

\begin{equation}
\label{h_frac_eqn}
\left(\frac{1}{\sqrt{2}}(\qi + \qk)\right)^\alpha \longleftrightarrow
\begin{bmatrix} 
\cos(\frac{\pi\alpha}{2}) + \frac{i}{\sqrt{2}} \sin(\frac{\pi\alpha}{2}) & \frac{i}{\sqrt{2}} \sin(\frac{\pi\alpha}{2}) \\
\frac{i}{\sqrt{2}} \sin(\frac{\pi\alpha}{2}) & \cos(\frac{\pi\alpha}{2}) - \frac{i}{\sqrt{2}} \sin(\frac{\pi\alpha}{2}) 
\end{bmatrix}
\longleftrightarrow H^\alpha
\end{equation}

Note that when $\alpha=1$, these matrices become $\begin{bmatrix} 0 & i \\ i & 0 \end{bmatrix}$ and $\frac{1}{\sqrt{2}}\begin{bmatrix} i & i \\ i & -i \end{bmatrix}$,
which are the standard $X$ and $H$ (Hadamard) rotation gates \cite[\S4.2]{nielsen2002quantum}, modulo a phase factor of $i$ to make the determinants equal to 1. 
This demonstrates the convenience of the quaternion formulation of Section \ref{quaternion-sec} --- while the fractional $X$-rotations of equation \ref{x_frac_eqn}
are standard in the literature, the fractional Hadamard gate of equation \ref{h_frac_eqn} is uncommon, but the quaternion recipe makes writing down an 
appropriate expression for fractional powers of $H$ straightforward. (This point is revisited in Section \ref{related-sec}.)

\begin{figure}[ht]
\caption{Results of Addition Circuits Fractional Powers of X Rotation and Hadamard Gate as Generators}
\label{result_plots}

\begin{tabular}{ccc}

\includegraphics[width=2in]{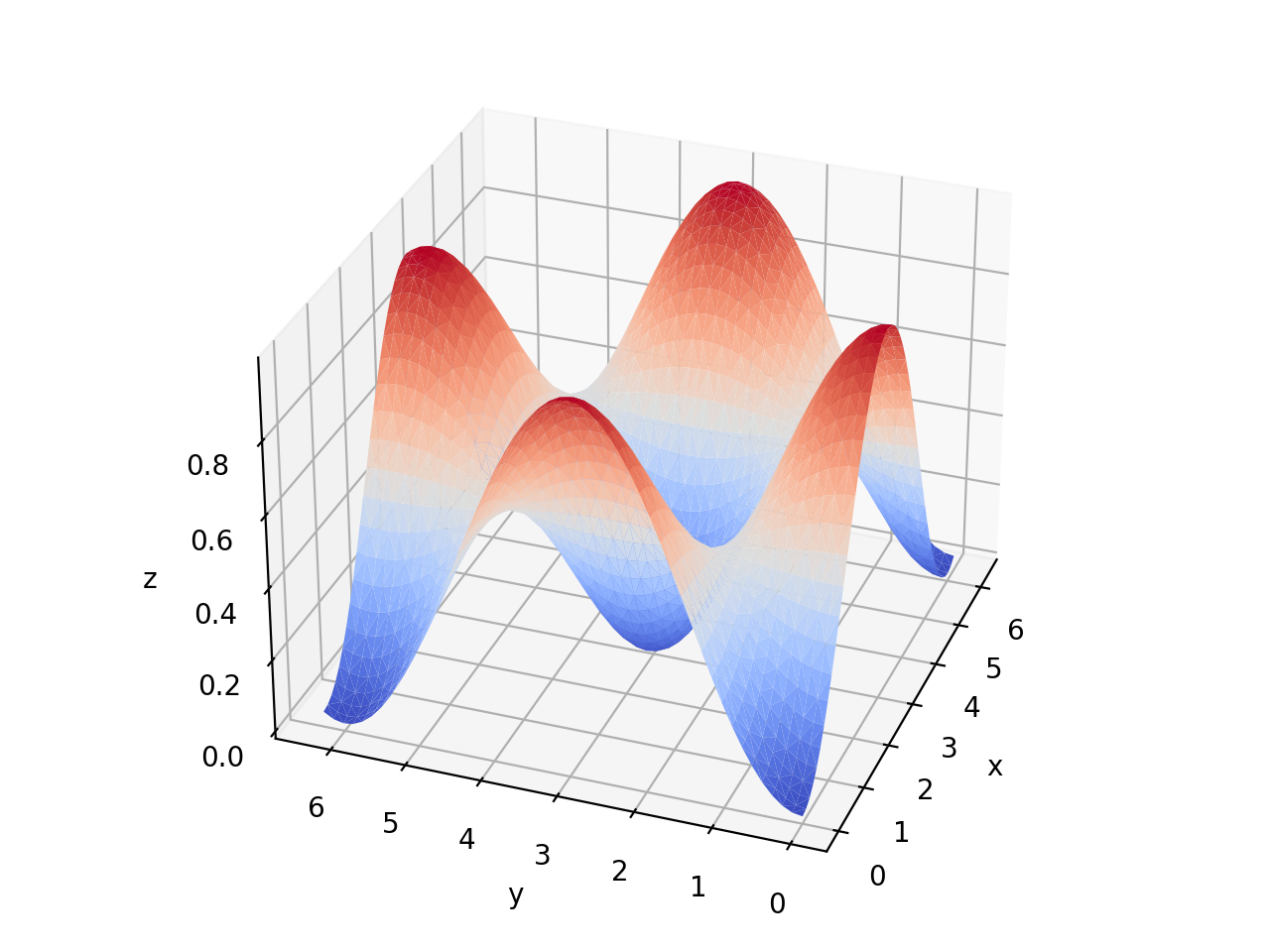} & 
\includegraphics[width=2in]{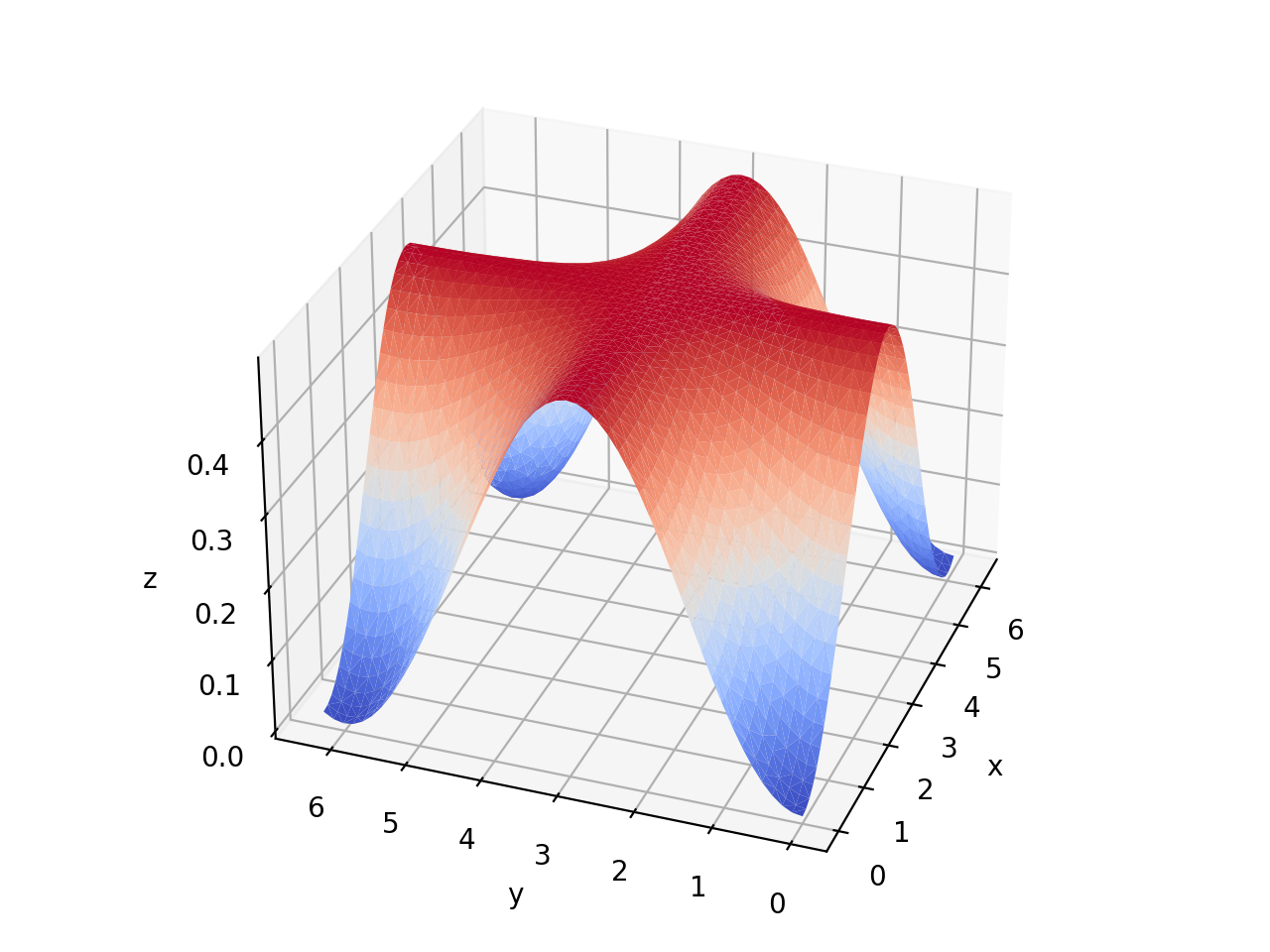} &
\includegraphics[width=2in]{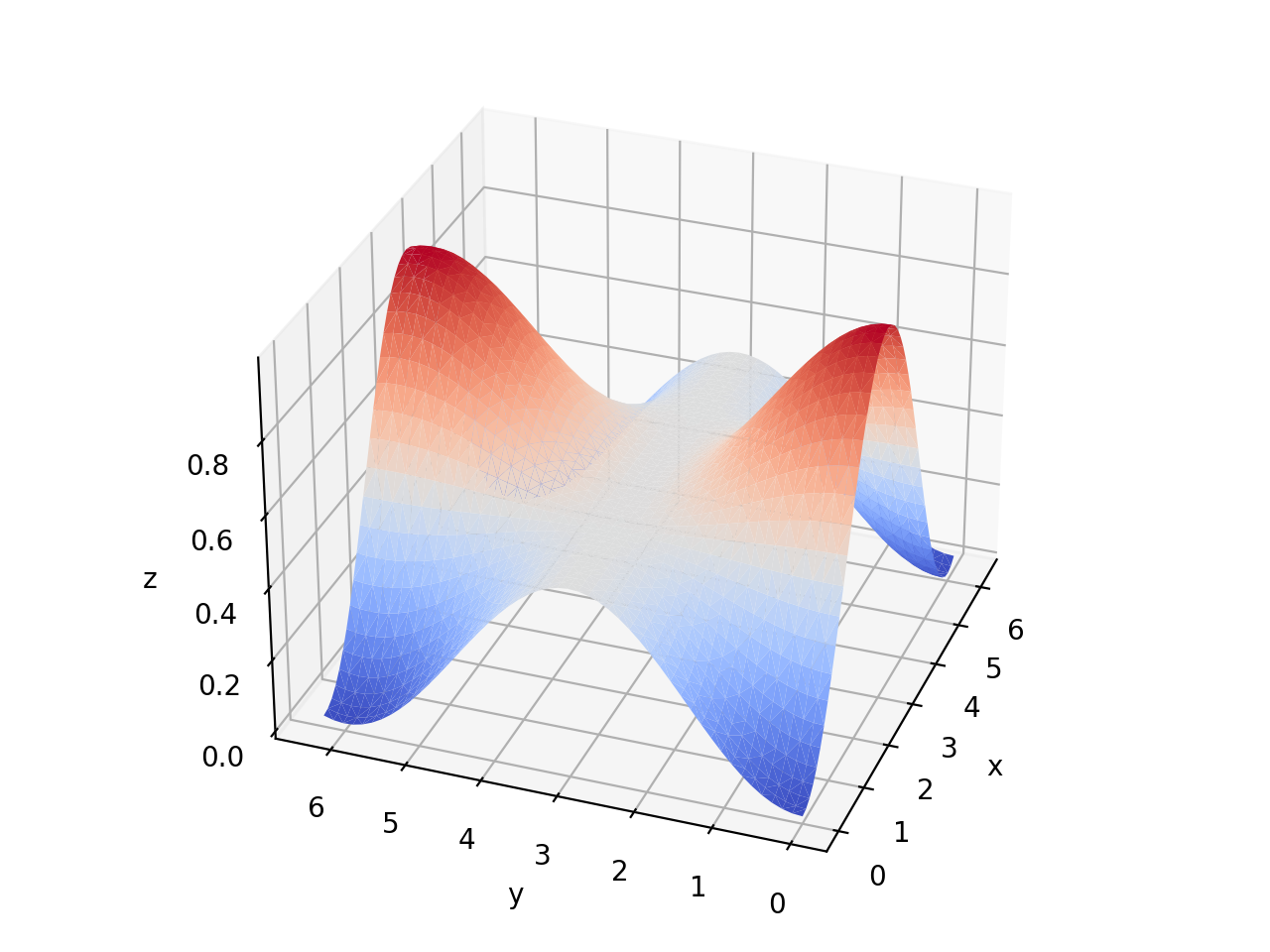} \\

Fractional X rotations & Fractional Hadamards & Mixture of X and Hadamard \\

\end{tabular}
\end{figure}

Example outcomes using combinations of these generators in parametrized circuits like those of Figure \ref{adder_circuit} are given in 
 in Figure \ref{result_plots}. Here the horizontal axes represent the angles $\theta$ and $\varphi$ through which the two summand qubits are
 rotated in Figure \ref{adder_circuit}, and the vertical axis represents the probability of measuring a $\ket{1}$ outcome in the sum qubit. 
 Around the origin, each combination 
is monotonically increasing as both the input angles increase, so for small angles, the result can be regarded as `nonlinear addition'. Larger angles exhibit periodic behavior as expected. 
The outcomes are different depending on the fractional generators used. 
With the fractional Hadamard generators, if either
summand is around the value $\pi$, the output is close to a maximum of $\frac{1}{2}$, which can be thought of as a continuous analogy to logical disjunction if the common
maximum value is rescaled to $1$. (Such a rescaling makes sense with this Hadamard algebra because it only takes values in $[0, \frac{1}{2}]$.)
The X rotation generators produce undulating waveforms with probabilities throughout the range $[0, 1]$, and less obviously, if either input is $\pi/2$, the outcome 
probabilities are always $50:50$. This is an intriguing alternative --- if the Hadamard disjunction shape can be summarized as ``whatever you say, I still say that one option is true'', 
the X rotation logic can be summarized as ``whatever you say, I still say that each option is equally likely''.
With a mixture of X rotation and Hadamard generators, we observe a combination of these behaviors. 

\subsection{Worked Example for Fractional X-Rotations}

This section works through the linear algebra of the circuit in Figure \ref{adder_circuit} with the generators set to be X rotations which correspond to the quaternion \qk.
For convenience, we can set $\theta = \pi\alpha$, so that $\qk^\alpha$ and $\mathrm{R_X}(\theta)$ are basically the same, and the equivalence in equation \ref{x_frac_eqn}
leads to the well-known form 
\[R_x(\theta) = \begin{bmatrix} \cos(\theta / 2) & -i\sin(\theta / 2) \\ -i\sin(\theta / 2) & \cos(\theta / 2) \end{bmatrix}\]


Consider the behavior of the quantum circuit in Figure \ref{adder_circuit} where operations on the top two qubits are $R_x(\theta)$ and $R_x(\varphi)$. 
For ease of notation, write $a = \cos(\theta/2)$, 
$b = -i\sin(\theta/2)$, $c = \cos(\varphi / 2)$, $d = -i\sin(\varphi/2)$, so that rotation matrices are just 
\begin{equation}
R_x(\theta) = \begin{bmatrix} a & b \\ b & a \end{bmatrix} \quad \mathrm{and} \quad 
R_x(\varphi) = \begin{bmatrix} c & d \\ d & c \end{bmatrix}
\label{rot_ab_eqn}
\end{equation} 

Now consider the combination of a single such rotation and a CNOT gate:

\begin{center}
\includegraphics[width=1.7in]{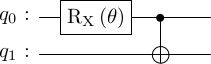}
\end{center}

\noindent
The action of this circuit on the 2-qubit basis states $\ket{00}$, $\ket{01}$, $\ket{10}$, $\ket{11}$ is given by the matrix
%
\[
\mathrm{CNOT} \circ (R_x(\theta) \otimes I_{2\times 2}) = 
\begin{bmatrix} 
a & 0 & b & 0 \\ 
0 & a & 0 & b \\ 
0 & b & 0 & a\\ 
b & 0 & a & 0 \\ 
\end{bmatrix}
\]
and the corresponding matrices for the 3-qubit versions of this component are shown in Table \ref{adder_components}. 
(The rows and columns of these matrices act upon the tensor product states in the big-endian order
$\ket{000}, \ket{001}, \ldots, \ket{111}$.)

\begin{table}[ht]
\caption{Circuits and unitary operator matrices for the given circuit components (zeros omitted)}
\label{adder_components}
\begin{center}
\begin{small}
\begin{tabular}{ccc}
\includegraphics[width=1.5in]{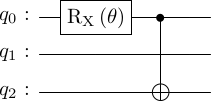} & &
$
\begin{bmatrix} 
a &&&& b \\
& a &&&& b \\
&& a &&&& b \\
&&& a &&&& b \\
&b &&&& a \\
b &&&& a \\
&&&b &&&& a \\
&&b &&&& a \\
\end{bmatrix}
$
\\
\bigskip 
\\
\includegraphics[width=1.5in]{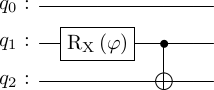} & &
$
\begin{bmatrix} 
c & & d \\
  & c & & d \\
  & d & & c \\
d & & c \\
&&&& c && d \\
&&&&& c && d \\
&&&&& d && c  \\
&&&& d && c\\
\end{bmatrix}
$
\\
\end{tabular}
\end{small}
\end{center}
\end{table}

The combined operation of the adder circuit of Figure \ref{adder_circuit} is given by the product of these two matrices. To skip tedious calculation, note
that we are only concerned with the action on the initial zero state $\ket{000}$, which corresponds to the vector $(1, 0, \ldots, 0)$. The first matrix in Table
\ref{adder_components} maps this vector to $(a, 0, 0, 0, 0, b, 0, 0)$, which the second matrix then maps to $(ac, 0, 0, ad, 0, bc, bd, 0)$. The nonzero entries
correspond to the states $\ket{000}$, $\ket{011}$, $\ket{101}$, $\ket{110}$, and the measurement is performed on the last qubit, so the probability of measuring
a $\ket{1}$ state for this qubit is given by the sum of the squares of the $\ket{011}$ and $\ket{101}$ components. It follows that
\begin{equation}
P(q_2 = \ket{1}) = |ad|^2 + |bc|^2 = \cos^2(\theta/2)\sin^2(\varphi/2) + \sin^2(\theta/2) \cos^2(\varphi/2).
\label{xrot_prob_result}
\end{equation}

This shows that the probability of observing a $\ket{1}$ in the target qubit is given by a nonlinear combination of the inputs
$\theta$ and $\varphi$, which near the origin is monotonically increasing in both $\theta$ and $\varphi$, and exhibits periodic behavior over wider ranges.
Less obviously, if either angle is equal to $\pi/2$, the outcome is a $50:50$ split. For example, if $\theta=\pi/2$, then $\cos^2(\theta)=\sin^2(\theta)=\frac{1}{2}$, 
so
\[ \cos^2(\theta/2)\sin^2(\varphi/2) + \sin^2(\theta/2) \cos^2(\varphi/2) = \frac{1}{2}(\sin^2(\varphi/2) + \cos^2(\varphi/2)) = \frac{1}{2}. \]

\subsection{Generalization to Other Quaternion Generators}

While the construction above is particularly straightforward for X-rotations, it generalizes to all unit quaternions relatively simply.
Recall from equation \ref{quat_to_su} that the quaternion $q = q_0 + q_1\qi + q_2\qj + q_3\qk$ 
corresponds to the matrix
\[\begin{bmatrix} 
q_0 + q_1i & q_2 + q_3i \\
-q_2 + q_3i & q_0 - q_1i \\
\end{bmatrix}
\]
so if we write $\alpha = q_0 + q_1i$ and $\beta = q_2 + q_3i$, then equation \ref{quat_to_su} leads to the correspondence
\begin{equation}  q =  q_0 + q_1\qi + q_2\qj + q_3\qk = \alpha + \beta \qj \longleftrightarrow A = \begin{bmatrix} \alpha & \beta \\ -\ov{\beta} & \ov{\alpha} \end{bmatrix}.
\end{equation}
Note that in this process, the quaternion $\qi$ is singled out and identified with the complex number $i$, which amounts to choosing a particular complex structure
that identifies the quaternions $\bb{H}$ with the complex vector space $\bb{C}^2$ \cite[\S1.1.2]{widdows2000quaternion}.
The matrix $A$ is only slightly more complicated than those representing X-rotations in equation \ref{rot_ab_eqn}. Relaxing the restriction that  
$A$ and $B$ should be fractional X-rotations, equation \ref{rot_ab_eqn} becomes
\begin{equation}
A = \begin{bmatrix}  \alpha & \beta \\ -\ov{\beta} & \ov{\alpha} \end{bmatrix} \quad \mathrm{and} \quad 
B = \begin{bmatrix}  \gamma & \delta \\ -\ov{\delta} & \ov{\gamma} \end{bmatrix}.
\label{qsum2_eqn}
\end{equation}
The matrix multiplication argument from the previous section can now be repeated, this time taking care the distinguish the top and bottom rows of the rotation
matrices. This leads to a generalization of the result $P(q_2 = \ket{1}) = |ad|^2 + |bc|^2$ in equation \ref{xrot_prob_result}, which is that
\[P(q_2 = \ket{1}) = |\alpha\ov{\delta}|^2 + |\ov{\beta}\gamma|^2. \]

This process also illustrates some of the useful flexibility of quaternion representations. Earlier in this paper, the calculation of powers and roots for
a particular quaternion could be done most simply by selecting a privileged square root of $-1$ for that quaternion, and using the corresponding polar representation
(equation \ref{quat_polar}). 
Now instead we want to reason about the combined effect of two quaternions, which is done easily by using a common square root of $-1$, and the corresponding
matrix representation.

The circuits and methodologies here extend to summing more than two inputs: this is done by adding more summand qubits and connecting them to the same
sum qubit using more CNOT gates. Initial computational experiments show that at least some of the properties found with two qubits persist with more qubits, including monotonic
increasing behavior in each variable near the origin, and sometimes if {\it any} input puts the output in a $50 : 50$ state, the result cannot be dislodged from there.
It is possible that an explicit expression like equation \ref{qsum2_eqn} can be found that generalizes to more input quaternions: this has not been done yet.
However, the next section demonstrates the practical effectiveness of such circuits in a machine learning task.

\section{System Application: Bag of Words Classification}
\label{classification-sec}

A `bag of words' classifier in natural language processing (NLP) is one that classifies texts or documents based purely on the words occurring in the document,
irrespective of the order they come in. For example, given the phrases ``horse chestnut'' and ``chestnut horse'', it has no way of knowing that the first is a kind of
tree and the second is a kind of animal --- all it knows is that {\it horse} makes the topic {\it animal} more likely and {\it chestnut} makes the topic {\it tree} more likely,
and it would score the phrases ``horse chestnut'' and ``chestnut horse'' identically. Though primitive, bag of words techniques were remarkably successful for decades
(especially in information retrieval \cite{widdows2004geometry}).

The process for using the adder components of Section \ref{addition-sec} to build a classifier was as follows:

\begin{itemize}
\item We are given a training corpus of (topic, sentence) pairs, with $T$ topics and a vocabulary of $W$ different words.
\item A circuit is created using $T(W+1)$ qubits. (This only works for very small training and test sets, hence this example would not yet scale to large datasets.)
\item The first $TW$ qubits store the (word, topic) scores, and the last $T$ qubits are the `topic counters' that accumulate a score for each topic at classification time.
\item During training, if the training phrase $(t, [w_1, \dots, w_n])$ is encountered, each of the qubits corresponding to the pair $(t, w_i)$ is incremented by a small angle.
\item During classification, for the test phrase $[w_1, \dots, w_n]$, each of the qubits $(t_j, w_i)$ is connected to the $t_j$ classification qubit using a CNOT gate, 
following the template component circuit in Figure \ref{adder_circuit}.
\item In this way, each of the topic classification qubits becomes an adder circuit scoring qubit for the word scores for that topic.
\item Each of the topic qubits is measured and (after an appropriate number of shots), the topic qubit with the highest probability of giving a $\ket{1}$ output is chosen.
\end{itemize}

\begin{figure}[ht]
\caption{Classification Circuit for Two Words and Two Topics}
\label{classifier_circuit}
\vspace{0.1in}
\includegraphics[width=3in]{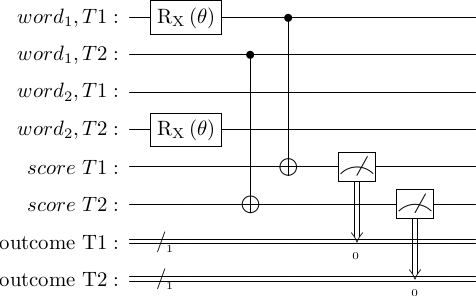}
\end{figure}

A minimal example circuit is shown in Figure \ref{classifier_circuit}. 
This circuit would arise from a training corpus with two training phrases, such as:
\begin{center}
\begin{tabular}{lcl}
I played football & $\rightarrow$ & \textsc{sport} \\
I played guitar & $\rightarrow$ & \textsc{music} \\
\end{tabular}
\end{center}
The implementation requires a classical preprocessing step which marks {\it football} as $\mathit{word}_1$, {\it guitar} as $\mathit{word}_2$,
\textsc{sport} as $T1$ and \textsc{music} as $T2$ (and ignores the other words since they are common to both topics). Then the quantum part of
the training process prepares the circuit in Figure \ref{classifier_circuit}, without the CNOT gates. For the quantum classification process, 
consider the phrase {\it ``I kicked the football''}. The classical register is used to recognize the word {\it football} as $\mathit{word}_1$, and 
so CNOT gates are added from the $word_1,T1$ and $word_1,T2$ qubits to the scoring qubits for each topic. Each of the topic qubits is measured (completing
the circuit outline in Figure \ref{classifier_circuit}), and the topic that gets the most $\ket{1}$ outcomes over a statistically significant number of shots
is chosen as the appropriate topic label.

This setup requires parametrization and tuning, and is still too basic to compete at scale with classifiers built
using large language models \cite[Ch 16]{geron2019hands}. However, it performs well
on a small dataset when compared to existing quantum alternatives. In particular, using a rotation angle of $\pi/24$ and a vocabulary reduced to 9 most salient words,
the classifier achieved 100\% accuracy on the test set from the lambeq datasets used in \cite{lorenz2021qnlp}, which contains
70 training phrases and 30 test phrases. This compares favorably with the 
accuracy of 83\% reported in \cite{lorenz2021qnlp}, which was achieved using just a 6 qubit register, but a much more linguistically 
sophisticated semantic modelling framework which requires a full grammatical parse tree as a classical preprocessing step. 
It should be noted that the dataset in question is artificially simple and was created specifically to enable experiments of this nature:
this more than anything else explains why it was possible to obtain 100\% accuracy.  
Our result here does not demonstrate that one method is better than the other in general --- rather, it showed that a simpler
model performed more effectively at this specific task at the cost of using more qubits, and that the components developed in this paper can 
be used to build circuits that produce successful results on real quantum hardware.

An obvious weakness of the classification circuit in Figure \ref{classifier_circuit} is that the CNOT gate is its own inverse,
so repeated operations cancel one another out rather than reinforcing one another. Given the test sentence {\it ``I like football, football is fun!''},
the repetition of the topical word {\it football} would cause the association with the correct topic to be forgotten rather than accentuated.
Another issue that remains to be investigated is automating the choice of a fractional power or rotation angle, which should vary 
between different words depending on their salience. There are also some positive observations to be made: in particular, the accuracy 
of the eventual classification decisions demonstrates that even with noisy intermediate scale quantum (NISQ) hardware, accurate
results can be obtained in cases where the application-level outcomes are suitably stable. In this demonstration case, the circuit
only needs to show that one topic is significantly more relevant than another, and this high-level decision was unaffected by small inaccuracies. 

\section{Related and Further Work}
\label{related-sec}

This paper has demonstrated a quaternionic method for finding arbitrary real powers of single-qubit gates, the use of these fractional powers in a 
simple addition component, and the use of several such components in a classifier circuit. This final section compares these approaches
to available alternatives.

The method of combining fractional rotations is different from the traditional ``quantum adder'' circuit proposed in \cite{feynman1985quantum}, 
which is for integer addition, and relies on using $n$-qubits to discretely represent $2^n$ bits. 
Optimizations obtained using the Fast Fourier Transform \citep{draper2000addition} speed 
up the computation, but still use the same discrete representation rather than using the continuous nature of vector coordinates or angles
to represent a continuum of real numbers. 
A natural question that arises is whether it is possible just to add two state vectors in a quantum computer, but this does not
work directly: the physical impossibility of a unitary transformation that takes $\ket{\Psi_1}$ and $\ket{\Psi_2}$ as inputs
and produces an output $\ket{\Psi_1} + \ket{\Psi_2}$ is discussed by \cite{alvarez2015forbidden}. (It is related to the no-cloning rule:
if $\ket{\Psi_1} = 0$, in many circumstances this would be equivalent to copying $\ket{\Psi_2}$ into the output state, which is impossible.)

The collection and combination of inputs from several components and outputting a score is ubiquitous in neural networks ---
different typesetting would make the circuit in Figure \ref{adder_circuit} look {\it exactly} like a network unit such as a perceptron
\cite[Ch 10]{geron2019hands}. The power of neural networks arises partly because the use of nonlinear activation functions enables 
them to approximate nonlinear as well as linear functions, and this is an acknowledged problem for quantum machine learning, because
unitary transformations are by definition linear  \cite[\S 5.4.1]{schuld2021machine}. This article has shown that, though matrices only
represent linear operators, when these are used to describe angles of unit quaternions, the algebraic operations on those angles are 
not at all linear. This motivates further work on testing these components on different machine learning problems, using a variety
of quaternion generators, to investigate which lead to the most successful nonlinear activation functions.

The equivalence between unit quaternions and special unitary matrices is well-known, and 
quaternions have been used to model operations on the Bloch sphere \citep{wharton2015unit}. This work uses the product of 
quaternions to model gate composition, but does not use quaternions to form powers and roots of gate operations.
Generators and roots of quantum gates are explored by \cite{muradian2005generators}. Their approach works for multi-qubit gates as well as single-qubit gates,
but only for gates $A$ such that $A^2=I$. An approximate method for calculating arbitrary real powers of square matrices is presented by \citet{higham2011schur}, 
who also review several other algorithms for this. This is a general solution to the matrix power problem, but relies on 
methods such as Schur decomposition and Pad\'e approximation, which are computationally detailed and less intuitively direct.
By contrast, the method developed here is particularly
simple for taking powers of the $2\times 2$ unitary matrices that represent single-qubit quantum gates. Due to its direct mathematical construction, the 
quaternionic method has particular geodesic properties, in the sense that it takes a smooth shortest-path through the quaternion group, and hence the
gate space $U(2)$ or the phase-free subgroup $SU(2)$, which may lend itself to exact and smooth implementation.

The geodesic property also helps to explain why the notion of `fractional rotations' does not work well with Euler angles. For example, it is 
well-known that a Hadamard gate can be decomposed as a $90^\circ$ rotation around the Y-axis, followed by a $180^\circ$ 
rotation around the X-axis \citep[\S 1.3.1]{nielsen2002quantum}. However, since rotations do not commute,
it does not follow that performing a $9^\circ$ Y-rotation followed
by an $18^\circ$ X-rotation 10 times in succession recovers the operation of a Hadamard gate --- a calculation shows
that this maps the state $\ket{0}$ to $(0.445-0.05i)\ket{0} + (0.632-0.632j)\ket{1}$ rather than the correct state $\frac{1}{\sqrt{2}}(\ket{0} + \ket{1})$.
The method of using quaternions instead of Euler angles could be described as finding a coordinate system in which dividing up rotations 
in such a way {\it does} work correctly. The drawbacks of Euler angles (such as coordinate singularities, gimbal lock, and cumbersome composition rules) 
are well-known \cite{diebel2006representing}, and such frustration has motivated the use of quaternions in areas of computer science including graphics \cite{mukundan2002quaternions}, 
3d simulation, and augmented reality   
\cite{kuipers1999quaternions} --- from this point of view, this paper extends some of the established benefits of quaternions over Euler angles to 
applications in quantum computing.

\section{Conclusion}

This paper demonstrated three main innovations:

\begin{enumerate}
\item The use of quaternion algrebra to represent fractional powers of any single-qubit gate.
\item The use of two CNOT gates to combine such fractions into a function of both inputs that can be chosen to be monotonically increasing over known intervals.
\item The arrangement of several such CNOT combinations to build a language topic classifier that performed perfectly accurately in a small experiment on quantum hardware. 
\end{enumerate}

The formulations developed here all have some novel aspects, though each of these advances can be derived quite easily from well-established mathematics, physics, and computer science. 
The small but successful classification example shows that the techniques can already be applied in a real application on quantum hardware.

\section{Acknowledgements}

Thanks for helpful discussions go particularly to Lazaro Calderin, Vandiver Chaplin, Jon Donovan, Chris Monroe, Daiwei Song, and Chase Zimmerman.

\bibliography{ionq}
\bibliographystyle{apalike}

\end{document}